\documentclass[journal=nalefd,manuscript=letter]{achemso}

\usepackage{graphicx}
\usepackage{epsfig}
\usepackage{epstopdf}
\usepackage{cleveref}
\usepackage{amsmath}
\usepackage{bm}
\usepackage{amssymb}

\usepackage{import}
\usepackage{color}
\usepackage{verbatim}
\usepackage{float}

\newcommand{\sign}{\mathop{\mathrm{sign}}\nolimits}
\renewcommand{\Im}{\mathop{\mathrm{Im}}\nolimits}
\newcommand{\onlinecite}[1]{\hspace{-1 ex} \nocite{#1}\citenum{#1}}

\title{Electrical switch to the resonant magneto-phonon effect in graphene}

\author{Przemyslaw Leszczynski}
\affiliation[LNCMI (CNRS, UJF, UPS, INSA), Grenoble, France]{LNCMI
(CNRS, UJF, UPS, INSA), BP 166, 38042 Grenoble Cedex 9, France}

\author{Zheng Han}
\affiliation[Institut N\'{e}el (CNRS-UJF-INP), Grenoble,
France]{Institut N\'{e}el, CNRS-UJF-INP, 38042 Grenoble Cedex 09,
France}

\author{Aurelien A. L. Nicolet}
\affiliation[LNCMI (CNRS, UJF, UPS, INSA), Grenoble, France]{LNCMI
(CNRS, UJF, UPS, INSA), BP 166, 38042 Grenoble Cedex 9, France}

\author{Benjamin A. Piot}
\affiliation[LNCMI (CNRS, UJF, UPS, INSA), Grenoble, France]{LNCMI
(CNRS, UJF, UPS, INSA), BP 166, 38042 Grenoble Cedex 9, France}

\author{Piotr Kossacki}
\affiliation[Institute of Experimental Physics, Warsaw,
Poland]{Institute of Experimental Physics, University of Warsaw,
Hoza 69, Warsaw 00-681, Poland}

\author{Milan Orlita}
\affiliation[LNCMI (CNRS, UJF, UPS, INSA), Grenoble, France]{LNCMI
(CNRS, UJF, UPS, INSA), BP 166, 38042 Grenoble Cedex 9, France}

\author{Vincent Bouchiat}
\affiliation[Institut N\'{e}el (CNRS-UJF-INP), Grenoble,
France]{Institut N\'{e}el, CNRS-UJF-INP, 38042 Grenoble Cedex 09,
France}

\author{Denis M. Basko}
\affiliation[LPMMC-CNRS, Grenoble, France]{Universit\'{e} Grenoble
1/CNRS, LPMMC UMR 5493, 25 rue des Martyrs, 38042 Grenoble,
France}

\author{Marek Potemski}
\affiliation[LNCMI (CNRS, UJF, UPS, INSA), Grenoble, France]{LNCMI
(CNRS, UJF, UPS, INSA), BP 166, 38042 Grenoble Cedex 9, France}

\author{Clement Faugeras}
\email{clement.faugeras@lncmi.cnrs.fr} \affiliation[LNCMI (CNRS,
UJF, UPS, INSA), Grenoble, France]{LNCMI (CNRS, UJF, UPS, INSA),
BP 166, 38042 Grenoble Cedex 9, France}

\begin{document}

\date{\today}

\begin{abstract}
We report a comprehensive study of the tuning with electric fields
of the resonant magneto-exciton optical phonon coupling in gated
graphene. For magnetic fields around $B \sim 25$~T which
correspond to the range of the fundamental magneto-phonon
resonance, the electron-phonon coupling can be switched on and off
by tuning the position of the Fermi level in order to Pauli block
the two fundamental inter Landau level excitations. The effects of
such a profound change in the electronic excitation spectrum are
traced through investigations of the optical phonon response in
polarization resolved magneto-Raman scattering experiments. We
report on the observation of a splitting of the phonon feature
with satellite peaks developing, at particular values of the
Landau level filling factor, on the low or on the high energy side
of the phonon, depending on the relative energy of the discrete
electronic excitation and of the optical phonon. Shifts of the
phonon energy as large as $\pm60$~cm$^{-1}$ are observed close to
the resonance. The intraband electronic excitation, the cyclotron
resonance, is shown to play a relevant role in the observed
spectral evolution of the phonon response.


\end{abstract}

\maketitle

In graphene, optical phonons from the Brillouin zone center
effectively couple to electronic excitations with the total
momentum $\mathbf{k}=0$. This coupling may lead to absorption of
phonons by electrons, which manifests itself as a broadening of
the $G$~peak in the Raman spectrum. Similar $\mathbf{k}=0$
electronic excitations determine the absorption of electromagnetic
radiation, proportional to the dissipative optical conductivity.
Despite the difference between the two families of electronic
excitations (phonons couple to valley-antisymmetric excitations,
while the optical conductivity is determined by valley-symmetric
excitations~\cite{Goerbig2007}), in many situations their spectra
can be assumed to be the same (unless some strong inter-valley
scattering processes are present). The low energy excitation
spectra of graphene can be modified by external means, for
instance by tuning the position of the Fermi level with an
electrostatic gate~\cite{Li2009n} or by applying an intense
magnetic field perpendicular to the plane of the graphene
crystal~\cite{Sadowski06,Jiang07}. The electron-phonon interaction
in graphene is particularly efficient for such optical-like
electronic excitations and, as a result, the effects related to
the electron-phonon interaction in graphene can be tuned
externally by modifying the electronic excitation spectrum.
Graphene hosts two Kohn anomalies at the $\Gamma$ and K points of
the Brillouin zone~\cite{Pisanec04}. The energies of optical
phonons at these two specific points are affected by the
electron-phonon interaction, at the level of 2-4$\%$. As a result,
the phonon energy as well as its line width, as seen through a
Raman scattering experiment, can be tuned externally by modifying
the electronic excitation spectrum, for instance, by changing the
position of the Fermi level~\cite{Pisana2007,Yan2007}, or by
applying a strong magnetic field~\cite{Ando07,Goerbig2007}.

An applied magnetic field (B) induces Landau quantization and
changes the continuous interband electronic excitation spectrum at
$B=0$ into a discrete excitation spectrum between highly
degenerated Landau levels of index $n$ with an energy that
increases with the magnetic field as
$E_n=\sign(n)\sqrt{2|n|}(\hbar{v}_F/l_B)$, where $v_F$ is the
Fermi velocity and $l_B=1/\sqrt{e|B|/\hbar}$ is the magnetic
length. The electronic excitations $L_{n,m}$ between Landau levels
of indices $n$ and $m$, relevant for the electron-phonon
interaction are those fulfilling the optical selection rule
$|n|-|m| = \pm 1$. In the following, we will refer to these
excitations as optical-like excitations. When these specific
excitations are tuned to the phonon energy by increasing the
strength of the magnetic field, the electron-phonon interaction
becomes resonant, between one discrete inter Landau level
electronic excitation and the phonon mode. This interaction
manifests itself through the magneto-phonon resonance, a series of
avoided crossings between the phonon mode and the electronic
excitation spectrum each time a $\Delta |n|=\pm 1$ inter Landau
excitation is tuned to the phonon
energy~\cite{Faugeras2009,Yan2010,Faugeras2011,Kuhne2012,Kossacki2012,Kim2013}.
Magneto-phonon resonance has also been observed in multi-layer
graphene specimens~\cite{Faugeras2012}, in bulk graphite for both
H and K point carriers~\cite{Kossacki2011,Kim2012} and is now a
tool to study the electron-phonon interaction and to perform the
Landau level spectroscopy of unknown systems. At the resonance,
the splitting energy depends on the effective oscillator strength
of the electronic excitation which is determined by the strength
of the magnetic field and by the occupancy of the initial and
final Landau levels implied in the excitation. Up to now, most
magneto-Raman scattering experiments have been performed on
graphene specimens with a fixed carrier density
$n_s$~\cite{Kossacki2012,Kim2013}. Recently, Remi et
al.~\cite{Remi2013} reported on unpolarized Raman scattering
experiments on a gated graphene sample, at $B=12.6$~T, in the
non-resonant regime. Here, we demonstrate experimentally that, by
changing the position of the Fermi level among the Landau levels
with electrostatic gating, it is possible in graphene in a
magnetic field, to externally switch on and off the resonant
electron-phonon interaction.

For high enough magnetic fields ($B>5$~T), all interband Landau
level excitations have an energy higher than that of the optical
phonon, except for the $L_{-1,0}$ and $L_{0,1}$ excitations, whose
transition frequency $T_0=\sqrt{2}\,v_F/l_B$ matches that of the
phonon at $B \sim 25$~T. The $L_{0,1}$ and $L_{-1,0}$ excitations,
active in distinct crossed circular polarization configurations,
when activated by adjusting the filling factor $\nu=h n_s/(eB)$
between $\nu=+6$ and $\nu=-6$, profoundly affect the phonon energy
through their resonant coupling to the phonon mode. This effect,
together with the coupling to the intraband cyclotron resonance
mode, can be traced by performing polarization resolved
magneto-Raman scattering experiments for different values of the
Fermi energy, the results of which are described in this paper.

\paragraph{\textbf{Experimental procedure}}

Graphene single grains on Cu were grown by chemical vapor
deposition (CVD). During the growth, temperature was kept at
1000$^\circ$C, while partial pressures for hydrogen and methane
were 25~$\mu$bar and 50~$\mu$bar, respectively. The growth was
stopped at 5 min, before graphene grains merge into a continuous
layer. The graphene grains were then transferred onto 285~nm
SiO$_2$/Si wafer with the PMMA-assisted method~\cite{Li2009b}.
Those randomly scattered graphene grains on SiO$_2$ were finally
contacted with arrays of long metallic leads (50~nm Au/5~nm Ti),
each lead serving as an individual ground. When patterned into a
Hall bar geometry, such samples show a typical electronic mobility
of $\sim 4000$~cm$^2$.(V.s)$^{-1}$. Polarization resolved Raman
scattering measurements have been performed with a home made
miniaturized optical bench~\cite{Faugeras2011} based on optical
fibers, lenses and band pass filters. The excitation laser at
$\lambda=514.5$~nm is delivered through a mono-mode optical fiber
of $\sim 5~\mu$m core, the scattered light is collected with a
50~$\mu$m optical fiber and analyzed by a 50~cm grating
spectrometer equipped with a nitrogen cooled charge coupled device
(CCD) camera. The excitation power was set to 4~mW focused on a
$\sim$~1~$\mu$m diameter spot. This system allows for the
measurement of the Raman scattering response of graphene at liquid
helium temperature and in magnetic fields up to 30~T. We have used
the crossed circular polarization configurations $\sigma \pm /
\sigma \mp$ (excitation light polarization/collection light
polarization), which select $\Delta|n|=\pm 1$ electronic
excitations and optical phonons at the $\Gamma$ point (G
band)~\cite{Kashuba2009,Kossacki2011} in graphene. Measurements
have been performed either at a fixed gate voltage while sweeping
the magnetic field, or at constant magnetic field while sweeping
the gate voltage.

\begin{figure*}[t!]
\begin{center}
\includegraphics[scale=0.8]{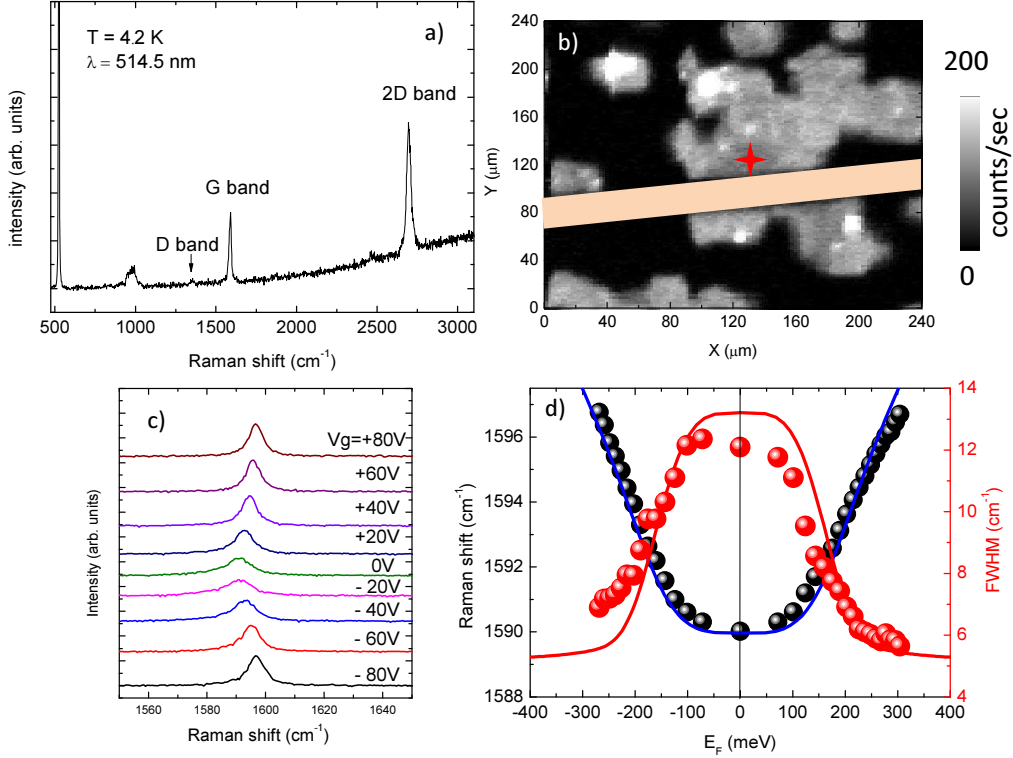}
\caption{a) Typical Raman scattering spectrum of the sample
measured at temperature 4.2~K with a 514.5~nm excitation. b)
Grayscale map of the G band feature intensity. The metallic gate
electrode is drawn in the middle of the flake and the red star is
the location at which measurements have been performed. c) Typical
spectra at $B=0$~T for different values of the gate voltage. d)
Evolution of the phonon energy (black points) and FWHM (red
points) as a function of the Fermi level. The blue (red) line is
the calculated evolution of the phonon energy (FWHM) according to
the model described in the text. A gaussian-type distribution of
carrier density with a standard deviation of
$\sigma=1.27\times10^{12}$~cm$^{-2}$ has been used (see text).}
\label{Fig1}
\end{center}
\end{figure*}

Figure~\ref{Fig1} a) shows a typical Raman scattering spectrum of
the graphene flake which has been investigated in magnetic fields.
The 2D band has a single Lorentzian shape, characteristic of a
graphene monolayer~\cite{Ferrari2013}. Figure~\ref{Fig1} b) shows
a spatial map of the G band intensity and helps to visualize the
shape of the flake together with the electrode. The bright spots
appearing in this figure correspond to bilayer graphene with a non
Bernal stacking, which are typical of CVD grown graphene. They are
characterized by a G band intensity 30 times enhanced in our
sample with respect to the G band feature at nearby locations, by
a single Lorentzian-shaped 2D band feature slightly blue shifted
and by an additional feature, the R band, observed in the present
case at 1490~cm$^{-1}$~\cite{Carozo2011,Kim2012,He2013} (see
supplementary information). Measurements were performed at the
location indicated by the red star in Figure~\ref{Fig1} b), close
to the electrode.

Tuning the position of the Fermi energy at low temperature leads
to a renormalization of the phonon energy and of its full width at
half maximum (FWHM)~\cite{Pisana2007,Yan2007} as a result of the
changes of the density of states at the Fermi energy and of the
gradual quenching of low energy electronic excitations due to
Pauli blocking. Typical Raman scattering spectra for different
values of the gate voltage are presented in Figure~\ref{Fig1} c).
The evolution of the phonon energy and FWHM as a function of the
Fermi energy, obtained experimentally, is shown in
Figure~\ref{Fig1} d) together with the result of the theoretical
modeling of this effect~\cite{Ando06}. In the modeling we used the
value $\omega_0=1588.7\:\mbox{cm}^{-1}$ for the phonon frequency
of the undoped sample at zero magnetic field, the dimensionless
electron-phonon coupling constant $\lambda=4.0\times 10^{-3}$, and
the Fermi velocity $v_{F}=1.08\times10^{6}$~m/s. The electronic
transitions are assumed to be homogeneously broadened with the
broadening parameter $\hbar\gamma=12$~meV. In addition, the
inhomogeneous spatial fluctuations of the electronic density are
accounted for by a convolution of the results obtained at a fixed
density, with the Gaussian distribution of densities,
characterized by a standard deviation of
$\sigma=1.27\times10^{12}\:\mbox{cm}^{-2}$. The Fermi velocity is
determined by the observation of the resonant coupling of the
$L_{0,1}$ inter Landau level excitation with the optical phonon at
$B\sim 25$~T (see supplementary materials). The Fermi energy $E_F$
was deduced from the applied gate voltage $V_g$ using the relation
$n_{s}=\alpha V_g$ where
$\alpha=7.56\times10^{10}$~cm$^{-2}$V$^{-1}$ for a 285~nm thick
SiO$_2$ layer and $E_{F}=\hbar v_{F} \sqrt{\pi n_s}$. A hysteresis
in the position of the Dirac point was observed between
consecutive gate voltage sweeps and shifts $\Delta V_{Dirac}$ up
to $\pm 8$~V have been corrected.

\paragraph{\textbf{Theoretical modeling of the Raman spectrum}}

The Raman spectrum is assumed to be proportional to the phonon
spectral function $-(1/\pi)\Im{D}_\pm(\omega)$, determined by
the retarded phonon propagator,
\begin{equation}
D_\pm(\omega)=\frac{2\omega_0}{\omega^2-\omega_0^2
-2\lambda\omega_0\,\Pi_\pm(\omega)}. \label{Spectralfunc}
\end{equation}
Here $\Pi_\pm(\omega)$ is the retarded polarization operator which
is diagonal in the circular basis, the two circular polarizations
being denoted by~$\pm$. The polarization operator can be
straightforwardly evaluated, at an arbitrary filling factor~$\nu$,
following the procedure described in
Ref.~[\onlinecite{Goerbig2007}, \onlinecite{Ando07}]. Keeping all
the non-resonant terms (some of which were omitted in
Refs.~[\onlinecite{Goerbig2007}, \onlinecite{Ando07}]), we obtain
\begin{eqnarray}
&&\Pi_\pm(\omega)=
\sum_{n=n_F}^\infty\frac{\omega(\omega+i\gamma)T_0^2/{T}_n}%
{(\omega+i\gamma)^2-{T}_n^2}
+\sum_{n=0}^{n_F-1}\frac{T_0^2}{{T}_n}+W_\nu,\\
&&W_{\nu>2}=\frac{T_0^2}2\left(\frac{f}{{T}_{n_F}}
-\frac{1-f}{{T}_{n_F-1}}
-\frac{1-f}{\omega_c^-}-\frac{f}{\omega_c^+}\right)+\nonumber\\
&&\qquad\quad{}+\frac{\omega{T}_0^2}2
\left[-\frac{f/{T}_{n_F}}{\omega+i\gamma\pm{T}_{n_F}}
+\frac{(1-f)/{T}_{n_F-1}}{\omega+i\gamma\mp{T}_{n_F-1}}
\right]+\nonumber\\
&&\qquad\quad{}+\frac{\omega{T}_0^2}2
\left[\frac{(1-f)/\omega_c^-}{\omega+i\gamma\mp\omega_c^-}
+\frac{f/\omega_c^+}{\omega+i\gamma\mp\omega_c^+}\right],\\
&&W_{-2<\nu<2}=\pm\frac{\omega{T}_0^2\,(2f-1)}%
{(\omega+i\gamma)^2-T_0^2},\\
&&W_{\nu<-2}=\frac{T_0^2}2
\left(\frac{1-f}{{T}_{n_F}}-\frac{f}{{T}_{n_F-1}}
-\frac{f}{\omega_c^-}-\frac{1-f}{\omega_c^+}\right)+\nonumber\\
&&\qquad\qquad{}+\frac{\omega{T}_0^2}2
\left[-\frac{(1-f)/{T}_{n_F}}{\omega+i\gamma\mp{T}_{n_F}}
+\frac{f/{T}_{n_F-1}}{\omega+i\gamma\pm{T}_{n_F-1}}
\right]+\nonumber\\ &&\qquad\qquad{}+\frac{\omega{T}_0^2}2
\left[\frac{f/\omega_c^-}{\omega+i\gamma\pm\omega_c^-}
+\frac{(1-f)/\omega_c^+}{\omega+i\gamma\pm\omega_c^+}\right].
\end{eqnarray}
Here we denoted by $n_F\geq{0}$ the (non-negative) index of the
Landau level which is partially filled (the integer part of
$|\nu|/4+1/2$), and by $f$ the average electronic occupation of
this partially filled level ($0\leq{f}<1$). We also introduced the
interband transition frequency $T_n=(\sqrt{n}+\sqrt{n+1})T_0$ and
the two intraband frequencies corresponding to the cyclotron
resonance, $\omega_c^+=(\sqrt{n_F+1}-\sqrt{n_F})T_0$ and
$\omega_c^-=(\sqrt{n_F}-\sqrt{n_F-1})T_0$. The electronic damping
$\gamma$ is introduced phenomenologically, keeping in mind that
$\Pi_\pm(\omega)$ should satisfy the general condition:
$\sign(\Im\Pi_\pm(\omega))=-\sign(\omega)$, and should be
continuous upon the change of the filling factor when
$n_F\to{n}_F+1$.

\paragraph{\textbf{Gate dependence close to the resonant condition}}

\begin{figure*}[t]
\begin{center}
\includegraphics[scale=0.8]{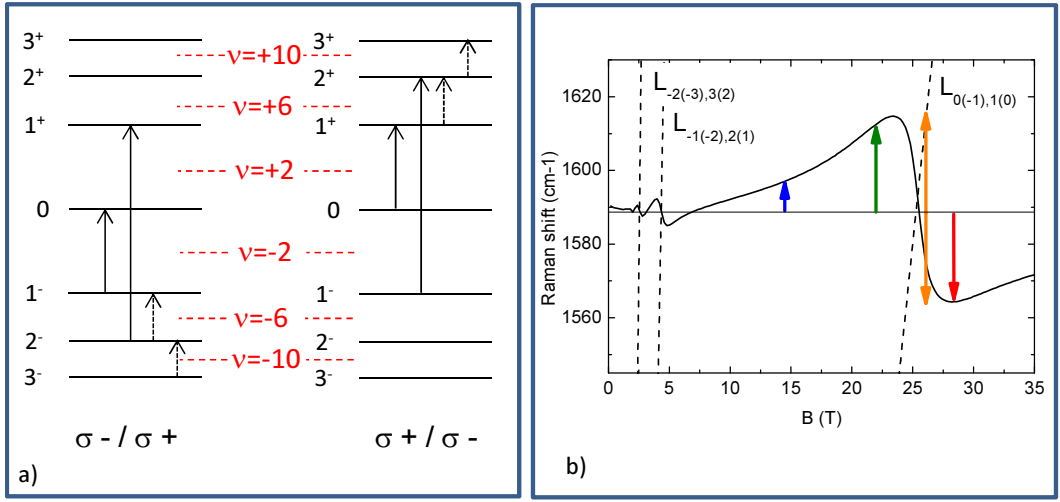}
\caption{a) Schematic of the Landau level structure in graphene at
finite magnetic field with relevant interband (solid arrows) and
intraband (dotted arrows) excitations in both polarization
configuration. b) Calculated evolution of the G band energy for
neutral graphene as a function of the magnetic field for $\sigma
+/\sigma -$ configuration. Dashed lines represent $\Delta |n|=\pm
1$ electronic excitations which couple to the phonon and colored
arrows indicate the magnetic fields at which the filling factor
has been experimentally varied. The length of the arrow indicate
the magnitude of the expected energy shift. The horizontal solid
line indicates the $B=0$ phonon energy.} \label{Fig2}
\end{center}
\end{figure*}

Figure~\ref{Fig2} a) shows a schematic of the 4-fold degenerate
Landau levels in graphene with an applied magnetic field together
with the interband (solid arrows) and intraband (dashed arrows)
excitations which are allowed in the two crossed circular
configurations $\sigma \pm / \sigma \mp$ and that can be probed
selectively with polarization resolved Raman scattering
techniques. Depending on the value of the filling factor, both
intraband and interband excitations can be active or quenched due
to Pauli blocking. For instance, the $L_{0,1}$ excitation is
allowed for $-2<\nu<+6$ and its effective oscillator strength has
the maximum at $\nu=+2$, while the $L_{-1,0}$ excitation is
allowed for $-6<\nu<+2$ and its effective oscillator strength is
the largest at $\nu=-2$. Intraband excitations are only active in
the $\sigma -/\sigma +$ configuration for $\nu>+2$ and in the
$\sigma +/\sigma -$ configuration for $\nu<-2$. Similar conditions
on the filling factor can be derived for these excitations and,
for instance, the $L_{1,2}$ excitation is allowed for $+2<\nu<+10$
and has a maximum effective oscillator strength at $\nu=+6$, when
the first Landau level is completely occupied and the second
Landau level is fully depleted. Figure~\ref{Fig2} b) shows the
evolution of the phonon energy for neutral graphene, which
displays the pronounced oscillations representative of the
magneto-phonon resonance, when the interband excitations are
tuned, one after another, in resonance with the phonon energy. We
have selected the values of the magnetic field, indicated by the
colored arrows in this figure, at which we have measured the
carrier density dependent magneto-Raman scattering response. At
$B=14$~T, the electron-phonon interaction is not resonant, but the
observed phonon energy is sensitive to either the $L_{0,1}$ or the
$L_{-1,0}$ excitations, which increase slightly the phonon energy
with respect to its $B=0$ energy, when they are
active~\cite{Remi2013}. These two particular electronic
excitations are tuned to the phonon energy at $B\sim 25$~T. Hence,
$B=22$, $26$ and $28$~T represent the different cases just below
the resonance, on the resonance and above the resonance,
respectively. The expected changes in the phonon energy, when the
$L_{0,1}$ or $L_{-1,0}$ excitations are turned on or off by
adjusting the carrier density, are completely different in these
three regimes: (i) below the resonance, the phonon energy is
expected to increase when $L_{0,1}$ or $L_{-1,0}$ is turned on,
(ii) close to the resonance, the phonon should split into two
components in clean enough systems, one with an increased energy
and the other one with a reduced energy, and, finally, (iii) above
the resonance, the phonon energy is expected to decrease.

\begin{figure*}[t]
\begin{center}
\includegraphics[scale=0.8]{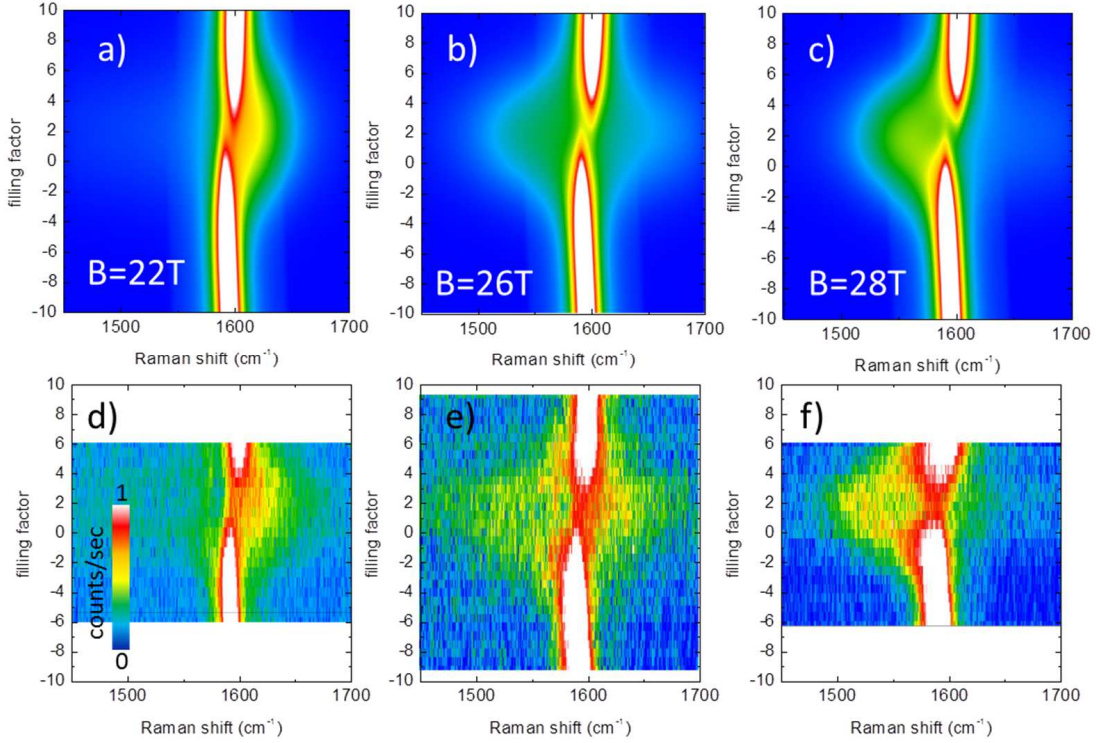}
\caption{a,b and c) False color map of the calculated phonon
spectral function in the $\sigma+/\sigma-$ polarization
configuration as a function of the filling factor for $B=22, 26 $
and $28$~T, respectively, and with a gaussian-type distribution of
carrier density with a standard deviation of
$1.27\times10^{12}$~cm$^{-2}$. d,e and f) False color map of the
experimental results in the same range of filling factors at
$B=22, 26$ and $28$~T respectively. The color scale has been
arranged in order to see the satellite peaks. The main phonon
feature is seen with a yield of 3.5 counts/sec.} \label{Fig3}
\end{center}
\end{figure*}

\begin{figure*}[t]
\begin{center}
\includegraphics[scale=0.8]{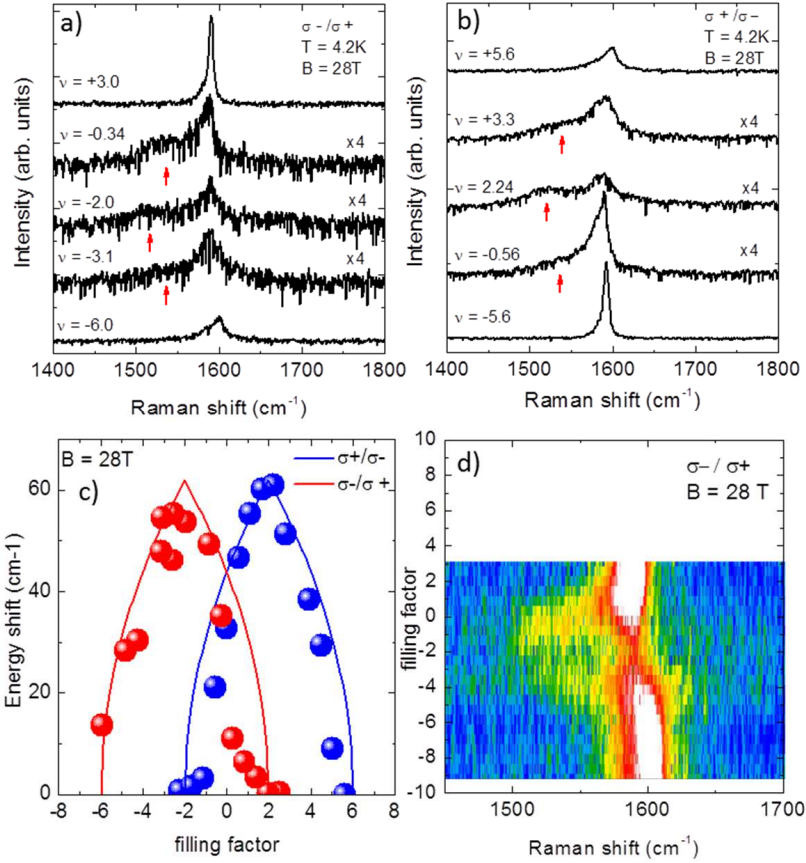}
\caption{a,b) Typical Raman scattering spectra measured at
$B=28$~T in the $\sigma-/\sigma+$ and in the $\sigma+/\sigma-$
polarization configuration respectively for different values of
the filling factor. Red arrows indicate the position of the
shifted phonon feature. c) Evolution of the phonon splitting
energy as a function of the filling factor in the
$\sigma-/\sigma+$ (red dots) and $\sigma+/\sigma-$ (blue dots)
together with the expected square-root dependence (solid lines).
d) False color map of the measured scattered intensity in the
$\sigma-/\sigma+$ polarization configuration as a function of the
filling factor at $B=28$~T.} \label{Fig4}
\end{center}
\end{figure*}

The results of such experiments performed in the $\sigma +/\sigma
-$ configuration are shown in Figure~\ref{Fig3} d,e and f) in the
form of false color maps of the scattered intensity in the range
of the optical phonon as a function of the filling factor for
$B=22, 26$ and $28$~T, respectively. Figure~\ref{Fig3} a,b and c)
show the result of the calculation of the spectral function as
defined in Equation~\ref{Spectralfunc}, for the same values of the
magnetic field, as a function of the filling factor. As before, we
have included inhomogeneous spatial fluctuations of the electronic
density by doing a convolution of the results obtained at a fixed
density, with the Gaussian distribution of densities,
characterized by a standard deviation of
$\sigma=1.27\times10^{12}\:\mbox{cm}^{-2}$. The observed shift of
the spectral weight of the phonon feature is representative of the
magneto-phonon resonance with shifts of the phonon energy as large
as $\pm 60$~cm$^{-1}$. As shown in Figure~\ref{Fig4}b), the
additional feature that appears when the $L_{0,1}$ excitation is
active in the $\sigma+/\sigma-$ configuration has a FWHM as large
as $\sim 50$~cm$^{-1}$, only exists when the filling factor is
between $\nu=-2$ and $\nu=+6$ and the observed shift is maximum
close to $\nu=+2$, when the effective oscillator strength of the
$L_{0,1}$ excitation is maximum. If the carrier density were
spatially homogeneous across the sample, a single component should
be observed for all values of the filling factor (see
supplementary information). Spatial inhomogeneities of the carrier
density at the nm-scale mix the Raman scattering response of
graphene locations with different apparent strength of
electron-phonon coupling. This leads to the observation of two
phonon features even when $-2<\nu<+6$. This behavior is reproduced
in the calculations by taking into account such inhomogeneities
(see supplementary information).

As can be seen in Figure~\ref{Fig4} a and d), in the
$\sigma-/\sigma+$ polarization configuration, a pronounced shift
of the phonon spectral weight is also observed but the amplitude
of the splitting is now maximum at $\nu=-2$, when the effective
oscillator strength of the $L_{-1,0}$ electronic excitation,
active in this configuration, is maximum. These polarization
resolved Raman scattering experiments unambiguously demonstrate
the coupling of electronic excitations and optical phonons in
graphene, both excitations having a similar angular momentum
$\pm1$. At $B=28$~T, we have determined the energy shift of the
phonon feature which is presented in Figure~\ref{Fig4} c) as a
function of the filling factor, together with the expected
square-root dependence (solid lines), as expected from
Ref.~[\onlinecite{Goerbig2007}]. This energy difference represents
half of the total energy split because at $B=28$~T, only the low
energy component of the coupled magneto-exciton-phonon mode is
observed. The overall agreement between these data and existing
theories is very good and such experiment offers a possibility to
trigger the resonant electron-phonon interaction by electrical
means and to gradually reach a strongly interacting regime.

\paragraph{Gate dependence at high filling factor values}

\begin{figure*}[t]
\begin{center}
\includegraphics[scale=0.8]{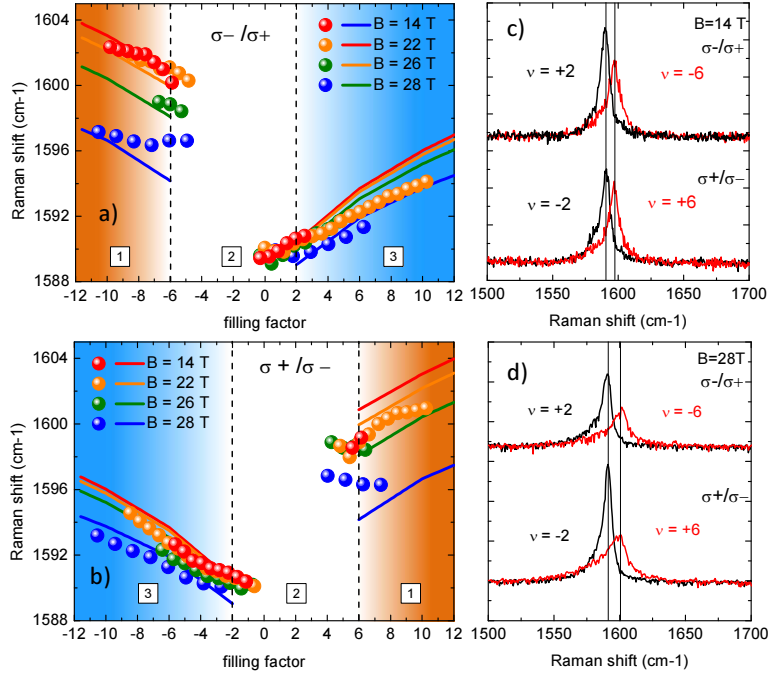}
\caption{a and b) Evolution of the phonon energy in the $\sigma
-/\sigma +$ and $\sigma +/\sigma -$ configurations respectively,
as a function of the filling factor for different values of the
magnetic field (points) together with theoretical expectations
including intra band excitations (solid lines). c,d) Typical
polarization resolved Raman scattering spectra for $\nu=+2$ and
$\nu=-6$ at $B=14$~T and $B=28$~T} \label{Fig5}
\end{center}
\end{figure*}

When the filling factor is tuned to high values so that the
$L_{0,1}$ or the $L_{-1,0}$ is turned off, the contribution of the
electron phonon interaction to the phonon energy is determined by
interband electronic excitations, involving Landau levels of high
index, with an energy much higher than the phonon energy, and by
intraband excitations. Gradually quenching such interband
excitations by increasing the Fermi energy (or the absolute value
of the filling factor) leads to an increase of the phonon energy.
In the range of magnetic fields addressed in this study, the
intraband excitations have an energy much lower than the phonon
energy (at $B=26$~T, the frequencies of the $L_{1(-2),2(-1)}$ and
$L_{2(-3),3(-2)}$ transitions are $T_1=670\:\mbox{cm}^{-1}$ and
$T_2=510\:\mbox{cm}^{-1}$, respectively) but they participate to
the phonon renormalization by slightly increasing its energy. This
effect is more pronounced as the magnetic field is increased, as
it is shown in Figure~\ref{Fig5} a,b) in both polarization
configurations. The solid lines in this figure represent the
expected phonon energy at different values of the magnetic field,
as a function of the filling factor. They were found by
calculating the positions of the poles of $D_\pm(\omega)$
(equation.~\ref{Spectralfunc}), in the complex plane $\omega$,
using the parameters corresponding to the experimental conditions
and presented in the preceding section. As can be seen in regions
3 of Figure~\ref{Fig5} a,b) when the intraband excitation is not
active (for $\nu>+2$ in $\sigma-/\sigma+$ configuration and for
$\nu<-2$ in $\sigma +/\sigma -$ configuration), the theoretical
expressions equation~\ref{Spectralfunc} describe quite well the
experimentally determined phonon energies, in terms of energy and
of its evolution when increasing the filling factor. In the
opposite range of filling factor, shown in parts 1 of
Figure~\ref{Fig5} a,b), the effect of intraband excitations is
clearly visible when comparing the spectra obtained at $\nu=+2$
and $\nu=-6$ in the $\sigma -/\sigma +$ configuration and spectra
at $\nu=-2$ and $\nu=+6$ in the $\sigma +/\sigma -$ (see
Figure~\ref{Fig5} c and d). Without considering intraband
excitations, these spectra should be identical because they
correspond to situations when only the fundamental interband
excitation is quenched while higher energy excitations are allowed
in both polarization configuration and have the same energy in the
first approximation. The spectra presented in Figure~\ref{Fig5}
c,d) clearly show that the phonon energy measured at these
particular values of the filling factor differ and that the energy
shift increases with increasing magnetic fields. These experiments
show however that there is a correspondence between spectra
measured in the $\sigma -/\sigma +$ at $\nu=-6$ and those measured
in the $\sigma +/\sigma -$ configuration at $\nu=+6$, which is a
consequence of the electron-hole symmetry. In these two
situations, the filling factor is such that the intraband
excitation ($L_{-2,-1}$ or $L_{1,2}$) has a maximized and
identical effective oscillator strength. The observed difference
in the phonon energy measured at $\nu=\pm 2$ and at $\nu=\mp 6$ is
a direct measurement of the coupling between the intraband
electronic excitation (cyclotron resonance) and the optical phonon
coupling, and is properly described by the expressions given in
equations~2-5.

\paragraph{\textbf{Conclusion}}
We have shown that the resonant interaction between
magneto-excitons and optical phonons in graphene can be switched
on and off by controlling the position of the Fermi level, and
hence, quenching or allowing the specific electronic excitations
through Pauli blocking. We have explored the three regimes for
which the electronic excitation lies below, is degenerated with or
is above the optical phonon mode. These three regimes have
distinct signatures when monitoring the position of the Fermi
level. In the resonant regime, satellite peaks appear when the
$L_{0(-1),1(0)}$ excitation is active and energy shifts of the
phonon feature up to 60 cm$^{-1}$ has been observed which evolves
with $\nu$ as expected theoretically. Experimental results in the
resonant regime, together with the effect of cyclotron resonance
are well reproduced by existing theories. Such experiments offer a
new insight into the electron-phonon interaction in graphene by
offering the possibility to continuously tune the electron-phonon
system from weakly to strongly interacting regime, and they could
be extended to multilayer graphene systems or to hybrid
graphene/2D materials heterostructures.

\begin{acknowledgement}
Part of this work has been supported by EuroMagNET II under the EU
contract number 228043, by the graphene flagship project and by
the European Research Council (ERC-2012-AdG-320590-MOMB)
\end{acknowledgement}

\providecommand*\mcitethebibliography{\thebibliography} \csname
@ifundefined\endcsname{endmcitethebibliography}
  {\let\endmcitethebibliography\endthebibliography}{}


\end{document}